\documentclass[aps,prb,twocolumn,superscriptaddress,floatfix,amsmath,longbibliography]{revtex4-2}
\usepackage{graphicx}
\usepackage{siunitx}
\usepackage{multirow}

\usepackage{hyperref}

\begin{document}

\title{Resonance energies and linewidths of Rydberg excitons in Cu$_2$O quantum wells}
\author{Niklas Scheuler}
\author{Patric Rommel}
\author{Jörg Main}
\email[Email: ]{main@itp1.uni-stuttgart.de}
\affiliation{Institut für Theoretische Physik 1, Universität
  Stuttgart, 70550 Stuttgart, Germany}
\author{Pavel A. Belov}
\affiliation{Institut für Physik, Universität Rostock,
  Albert-Einstein-Straße 23-24, 18059 Rostock, Germany}
	
\date{\today}

\begin{abstract}
Rydberg excitons are the solid-state analog of Rydberg atoms and can,
e.g., for cuprous oxide, easily reach a large size in the region of
$\mu$m for principal quantum numbers up to $n=25$. The fabrication of
quantum well-like structures in the crystal leads to quantum
confinement effects and opens the possibility to study a  crossover
from three-dimensional to two-dimensional  excitons. For small widths
of the quantum well (QW) there are several well separated Rydberg
series between various scattering thresholds leading to the occurrence
of electron-hole resonances with finite lifetimes above the lowest
threshold. By application of the stabilization method to the
parametric dependencies of the real-valued eigenvalues of the original
three-dimensional Schr\"{o}dinger equation we calculate the resonance
energies and linewidths for Rydberg excitons in QWs in regimes where a
perturbative treatment is impossible. The positions and finite
linewidths of resonances at energies above the third threshold are
compared with the complex resonance energies obtained within the
framework of the complex-coordinate-rotation technique. The excellent
agreement between the results demonstrates the validity of both
methods for intermediate sizes of the QW-like structures, and thus for
arbitrary widths.
\end{abstract}
	
\maketitle

\section{Introduction}
\label{sec:intro}
Excitons are  electron-hole bound states in semiconductors~\cite{Rashba}.
The exciton states in bulk cuprous oxide (Cu$_{2}$O) crystals are featured
by relatively large binding energies as well as by unique radiative characteristics.
The Rydberg states of excitons are, in turn, remarkable due to large
spatial extent of the wave function and thus high sensitivity to the
external fields~\cite{Schweiner17a} and surrounding
quasiparticles~\cite{Stolz2019}.
In cuprous oxide, the typical size of the Rydberg exciton
is in the order of $\mu$m for the principal quantum number of about
$n=25$~\cite{Kaz14,Thewes15,Schweiner16b}.
The binding energies of the Rydberg states are less than 10~meV,
though the tunability of the emitting light wavelengths makes them
convenient for practical applications.
This flexibility can be achieved by applying external
fields~\cite{Farenbruch2020,Zielinski2020} or by fabrication of
low-dimensional structures~\cite{Naka2018,Konzelmann2020,Belov2019,Belov2022}.
The latter can significantly change the energy spectrum of
electron-hole pairs: in addition to features of the band structure,
each quantum-confinement subband produces a proper Rydberg
electron-hole series~\cite{Belov2019}.
As a result, due to the Coulomb coupling of the upper subbands to the
continuum of lower subbands, the electron-hole resonant states appear
in the spectrum above the electron-hole scattering threshold.

Samples of  cuprous oxide crystals of the size of order of $\mu$m
are already produced, however, up to date, their quality allows one to
observe only several lower exciton states~\cite{Naka2018}.
Nevertheless, the technology is permanently developing and the low-dimensional
samples of the size of hundreds nm with high radiative properties are
expected to appear soon~\cite{Konzelmann2020}.
Very recently the fabrication of thin films with widths of about
20-30\,nm has been reported by Awal et al.~\cite{Awal2024}.
In this sense, the ways to decrease the nonradiative broadening of the
excited electron-hole states are especially
important~\cite{Poltavtsev2014}.

The simplest and most studied low-dimensional structure is a quantum
well (QW)~\cite{ivchenko}.
The exciton states in such a semiconductor
structure have been actively studied for many years~\cite{klingshirn2007semiconductor}.
For example, there are recent studies of excitons in GaAs-, CdTe-, and
GaN-based heterostructures~\cite{Laird2022,Kotova2023,Chiaruttini2021}.
However, the electron-hole resonances in such structures have been
studied less intensively, mainly because of the square-unintegrable
nature of the resonance wave function and resulting additional
nonradiative linewidth broadening.
Moreover, the results of earlier theoretical works on the broadening
of excitons in GaAs-based QWs due to coupling to the
continuum~\cite{Broido1988,Pasquarello1991} contradict more recent
analytical estimates~\cite{Schmelcher2005}.
The first numerical investigations of electron-impurity and
electron-hole resonant states using the finite-difference approach
combined with the complex-coordinate-rotation method~\cite{Rei82,Ho83,Moi98}
have been done in Refs.~\cite{Belov2022,BelResQW}.
For the bulk Cu$_{2}$O, there are also only a few recent studies
of resonance linewidth broadenings~\cite{Rommel2020,PhysRevB.104.085204}.
There is still no systematic study of the electron-hole resonances in QW-like structures,
neither for the well-known GaAs-based structures, nor for the recently
appearing cuprous oxide ones.

For the electron-hole pairs in GaAs-based structures, where the
electron is much lighter than the hole, the problem can be reduced to
a lower-dimensional model of an electron-impurity in a QW~\cite{Belov2022}.
By contrast, comparable effective masses of the electrons and holes in
cuprous oxide complicate the resonance calculation~\cite{Schweiner16b}.
Furthermore, the perturbative solution of the Schr\"{o}dinger equation
for Rydberg excitons is only possible  for the two limiting models of
strong (narrow QW) and weak confinement (wide QWs, approaching the
bulk crystal)~\cite{Belov2023}.
However, for the model of intermediate QW widths, a perturbative
treatment is inapplicable and one has to solve the original
three-dimensional equation numerically.
The numerically obtained linewidth broadenings as well as reproduced
absorption spectra based on mechanisms of exciton coupling to the
continuum and to the phonon background can test the applicability of
the Fano theory of resonances~\cite{Fano1961}.
Moreover, the different parity of the subband eigenfunctions leads to
a trivial example of the bound states in the
continuum~\cite{Stillinger1975,Moiseyev1976,Hsu2016} and may further
imply an existence of upper-lying states with negligible nonradiative
broadening.

In this Paper, we  calculate and identify the energies and the
linewidths of electron-hole resonances in semiconductor
heterostructures with intermediate QW widths for which the
perturbative treatment is impossible.
By our study, we show that the resonance parameters of Rydberg
excitons in the continuum background can be readily obtained for
arbitrary widths and relatively high energies.
We use a hydrogenlike two-band model of the electron-hole pair in a QW
structure~\cite{ivchenko} with the cuprous oxide material
parameters~\cite{Schweiner17a}, thus simulating the Cu$_2$O thin
film sandwiched by vacuum or air.
Note that the hydrogenlike model provides qualitatively good results
to describe the exciton Rybgerg series in the bulk up to
small deviations caused by the band structure~\cite{Kaz14}.
Here, we disregard features of the band structure, image
charges~\cite{Tran1990} and finite potential in the substrate, as well as
other effects which, for example, have been extensively employed to
model the electron-hole bound states in GaAs-based QWs~\cite{Belov2017,Belov2019}.
Despite the above-mentioned assumptions, such a model leads to the
three-dimensional Schr\"odinger equation~\cite{Khramtsov2016},
which produces the energy spectrum with quantum-confinement subbands
and many different branches of the continua.
Due to the Coulomb coupling of the upper quantum-confinement subbands
to the continuum of lower ones, multiple electron-hole resonant
states appear above the exciton states, namely above the scattering
threshold $E_{\mathrm{e}1}+E_{\mathrm{h}1}$, where $E_{\mathrm{e}1/\mathrm{h}1}$
is the lowest energy of the electron/hole in the QW.
We show that the parity dependence of the eigenfunctions gives rise
to the Rydberg series of bound states in the continuum~\cite{Hsu2016}.

To identify resonances above the lowest scattering threshold,
we take advantage of the stabilization method~\cite{Hazi1970,Mandelshtam1993,Mandelshtam1994}
and compare the results with data obtained by the
complex-coordinate-rotation technique.
The complex-coordinate rotation is a general and well established
method for the computation of resonances in open quantum systems.
The method has already been used to investigate stationary systems,
e.g., the resonant states of the hydrogen atom in external
fields~\cite{Delande91,Mai92}, the helium
atom~\cite{PhysRevA.56.1855}, bulk and QW
excitons~\cite{Zielinski2020,Rommel2020,BelResQW}, as well as
time-dependent scattering problems~\cite{Vinitsky,BelovMicro2023}.
All above-cited theoretical works on semiconductor physics use
this method as a reliable tool for obtaining the resonance energies
and the linewidth broadenings.
Although this approach is quite precise, it usually complicates the
Hamiltonian by introducing the artificial complex-rotational
parameters, and, thus, leads to a solution of the non-Hermitian
eigenvalue problem.
An alternative technique which allows one to estimate the resonance
parameters without making a complex rotation is the stabilization method.
It has already been applied in a variety of
works~\cite{Austin1985,Moi1994,Zhang2008,Moi2022,Hiyama2022}.
It solely implies rigid-wall boundary conditions at some distance from
the interaction domain. Taking this distance as a parameter, the
stabilization method is based on the observation of the parametric
dependence of the energies of the discretized continuum.
The avoided crossings of the discretized-continuum energy levels in
the vicinity of the proper resonant energies allow one to estimate the
linewidth broadenings of the resonances.

In our calculations within the stabilization method, the B-spline
basis is employed to precisely compute the energy levels of the
electron-hole pairs as well as their crossings and avoided crossings
as a function of the QW width.
The B-spline basis representation of the wave function allows for the
fast and accurate solution of the three-dimensional Hermitian
eigenvalue problem.
Besides the high accuracy of the results produced  by the high-order
B-spline calculations, the minimal support of the B-splines leads to
minimal overlaps and thus to the band structure of the matrices of the
corresponding generalized eigenvalue problem.
Moreover, the obtained parametric dependence of the energies on the QW width
makes it possible to accurately determine the linewidths of the
electron-hole resonant states by the stabilization method.
Comparison of the results computed by the stabilization method
with ones obtained from the non-Hermitian eigenvalue problem in the
framework of the complex-coordinate-rotation technique shows an
excellent agreement.
It justifies the validity of both methods for the intermediate sizes
of the QW-like structures, and thus for arbitrary widths.
This also paves the way for more detailed calculations and facilitates
further analysis of the experimental data~\cite{Naka2018}.

\section{Theory and methods}
We now present and discuss the Hamiltonian of the exciton in a QW, its
energy spectrum, the B-spline basis for the efficient numerical
solution of the Schr\"odinger equation, and two different methods for
the computation of resonance energies and linewidths.

\subsection{Excitons in cuprous oxide QWs}
The two-band model for the electron-hole pair in a cuprous oxide QW is
given as
\begin{align}
  \label{eq:origin_Hamilton}
	H &= E_{\mathrm{g}} - \frac{\hbar^2}{2m_{\mathrm{e}}}\Delta_{\mathrm{e}}
	- \frac{\hbar^2}{2m_{\mathrm{h}}}\Delta_{\mathrm{h}} \nonumber\\
	&- \frac{1}{4\pi\varepsilon_0\varepsilon}
	\frac{{e}^2}{|\boldsymbol{r}_{\mathrm{e}}-\boldsymbol{r}_{\mathrm{h}}|}
	+ V_{\mathrm{e}}(z_{\mathrm{e}})
	+ V_{\mathrm{h}}(z_{\mathrm{h}}) \, ,
\end{align}
with
\begin{equation}
	V_{\mathrm{e,h}}(z_{\mathrm{e,h}}) = \left\{
	\begin{array}{ll} 0,& \mathrm{if}\;|z_{\mathrm{e,h}}| < L/2\\
		\infty,& \mathrm{if}\;|z_{\mathrm{e,h}}| > L/2 \end{array} \right. .
\end{equation}
Here, $m_{\mathrm{e}}=0.99\,m_0$ and $m_{\mathrm{h}}=0.69\,m_0$ are the
effective masses of the electron (e) and the hole (h), respectively,
$E_{\mathrm{g}}=2.17208\,$eV is the band gap energy, ${e}$ is the electron
charge, $\varepsilon_0$ is the electric constant, $\varepsilon=7.5$ is
the dielectric constant, and $L$ is the width of the QW.
The infinite potential in the substrate is an approximation, however,
such an infinite-barrier potential can be a reasonable model for a thin
cuprous oxide film surrounded by vacuum or air \cite{Konzelmann2020}.
We separate the center-of-mass motion in the QW plane and use
polar coordinates $(\rho,\phi)$ to describe the relative motion.
The rotational symmetry around the $z$-axis implies the conservation
of the $z$ component of the angular momentum with quantum number $m$,
and we arrive at the three-dimensional equation
\begin{equation}\label{eq:schrodinger_eq}
  H\Psi(\rho,z_{\mathrm{e}},z_{\mathrm{h}}) = E \Psi(\rho,z_{\mathrm{e}},z_{\mathrm{h}}),
\end{equation}
with the Hamiltonian
\begin{align}
  \label{eq:Hamiltonian_in_com_coordinates}
	H &= E_{\mathrm{g}} - \frac{\hbar^2}{2\mu}\left(\frac{\partial^2}{\partial\rho^2}
	+ \frac{1}{\rho}\frac{\partial}{\partial\rho}-\frac{m^2}{\rho^2}\right) 
	- \frac{\hbar^2}{2m_{\mathrm{e}}}\frac{\partial^2}{\partial z_{\mathrm{e}}^2}
	- \frac{\hbar^2}{2m_{\mathrm{h}}}\frac{\partial^2}{\partial z_{\mathrm{h}}^2} \nonumber\\
	&- \frac{1}{4\pi\varepsilon_0\varepsilon}
	\frac{e^2}{\sqrt{\rho^2+(z_{\mathrm{e}}-z_{\mathrm{h}})^2}}
	+ V_{\mathrm{e}}(z_{\mathrm{e}}) + V_{\mathrm{h}}(z_{\mathrm{h}})\, .
\end{align}
Here, $\mu$ is the reduced mass, and the angular momentum quantum
number $m=0,\pm 1, \pm 2,\ldots$ is a good quantum number.
Besides $m$ there is one additional exact quantum number, i.e., the
parity $\pi_{z}=\pi_{z_{e}}\pi_{z_{h}}=\pm 1$, which is related to the
simultaneous exchange of $z_{\mathrm{e}}\to -z_{\mathrm{e}}$ and
$z_{\mathrm{h}}\to -z_{\mathrm{h}}$.

\subsection{Energy spectrum}
The Hamiltonian~\eqref{eq:Hamiltonian_in_com_coordinates} defines the
energy spectrum of the electron-hole  bound states and resonances in the QW.
The bound states correspond to the square-integrable solutions of the
Schr\"odinger equation, whereas the resonances are associated to the
square-unintegrable ones.
A resonant state can be considered as a quasi-bound state with a
finite lifetime due to an exponential decay, which can be described as
$\Psi(t) = \exp(-iEt/\hbar)\Psi(0)$
with a complex energy~\cite{Landau,Moiseyev1978}
\begin{equation}
  \label{eq:complex_energy}
  E = E_\mathrm{res} -i\frac{\Gamma}{2}  \, ,
\end{equation}
where $E_{\mathrm{res}}$ is the resonance position and $\Gamma>0$ is
the linewidth.
It means that the resonance wave function does not exponentially
decrease as $\rho \to \infty$, and thus these states cannot be
determined in terms of the Hermitian quantum theory~\cite{Moi98}.

Each disjoint subsystem (the electron or the hole in the QW)
of the Hamiltonian~\eqref{eq:Hamiltonian_in_com_coordinates}
produces a series of quantum-confinement energies $E_{\mathrm{e}i/\mathrm{h}j}(L)$
with $i,j=1,\ldots,\infty$.
Moreover, the Coulomb attraction leads to Rydberg energy series below
each value of the sum
\begin{equation}
  E_{i,j}(L) \equiv E_{\mathrm{e}i}(L) + E_{\mathrm{h}j}(L)
  = \frac{\hbar^2(i\pi)^2}{2m_{\mathrm{e}}L^2}
  + \frac{\hbar^2(j\pi)^2}{2m_{\mathrm{h}}L^2} \, .
\label{eq:thresholds}
\end{equation}
For a particular $L$, each such a value appears to be a scattering
threshold, giving rise to a certain branch of the continuum.
In case of  strong confinement, exciton states and electron-hole
resonances with the respective parity are located below the thresholds.
In this limit, the exciton wave function is factorized
$\Psi_{nij}(\rho,z_{\mathrm{e}},z_{\mathrm{h}})=R^{2D}_{n}(\rho) \psi_{\mathrm{e}i}(z_{\mathrm{e}}) \psi_{\mathrm{h}j}(z_{\mathrm{h}})$,
where $R^{2D}_{n}(\rho)$, $n=1,\ldots,\infty$ are the 2D Coulomb
radial eigenfunctions and
$\psi_{\mathrm{e}i/\mathrm{h}j}(z_{\mathrm{e}/\mathrm{h}})$ are the
quantum-confinement eigenmodes~\cite{Belov2019}.
The quantum numbers $i$, $j$ and the principal quantum number $n$ are the good ones.
Note that the Rydberg formula reads $E_n=-E_{\mathrm{Ryd}}/(n-1/2)^2$ in two dimensions.
For weaker confinement, the wave function is no longer factorized and
the quantum numbers of the states can only be associated
approximately.
The resonances with different parity can strongly overlap.
Below the lowest threshold, $E_{1,1}=E_{\mathrm{e1}}+E_{\mathrm{h1}}$,
there are the bound states.
Above this threshold the resonant states appear in the continuum
background~\cite{Belov2019}, except for odd parity bound
states, which, for $m_{e}>m_{h}$, appear up to the first odd parity
threshold $E_{2,1}=E_{\mathrm{e2}}+E_{\mathrm{h1}}$.
The decay of these odd parity bound states in the continuum
background of the even parity states is forbidden, i.e., the widths
are exactly $\Gamma=0$~\cite{Hsu2016}.

The energies of the states defined by
Eq.~\eqref{eq:Hamiltonian_in_com_coordinates} can be obtained
analytically only in the limiting cases of strong confinement or weak
confinement, when the Coulomb potential or the effect of QW barriers
can, respectively, be treated as a small perturbation~\cite{Belov2023}.
For arbitrary QW width $L$, Eq.~\eqref{eq:schrodinger_eq} can only be
solved numerically~\cite{Khramtsov2016}.
Moreover, the efficient numerical approach allows one to apply
the stabilization method~\cite{Hazi1970,Mandelshtam1993,Mandelshtam1994}
or the complex-coordinate-rotation method~\cite{Rei82,Ho83}
to go beyond Hermitian physics and to precisely estimate not only the
resonance energies, but also the linewidths.

\subsection{Computation of the energy spectrum}
The Schr\"odinger equation~\eqref{eq:schrodinger_eq} was already
treated numerically using the finite-difference and basis-expansion
approaches \cite{Khramtsov2016,JWilkes2016, Ziemkiewicz2021}.
These methods are appropriate for the bound state
calculations of lower exciton states.
However, for a precise determination of
the Rydberg energies or resonances associated to upper scattering
thresholds one has to use fine grids over a  broad calculation domain
or take many basis functions.
Moreover, the integration of the quantum-confinement basis functions
having nonzero support over the whole domain is usually also quite
time-consuming.

To overcome these computational issues and accurately estimate the
resonance linewidths, we use the expansion of the wave function over a
basis of B-splines~\cite{DeBoor}.
The B-splines $B^{k}_{i}(x)$, $i=1,\ldots,n$ are piecewise polynomials
of degree $k-1$, which can be generated by recursive formulas with
knots $t_{i}$~\cite{Bachau2001}.
For example, for $k=1$ the B-splines are piecewise-constant, for $k=2$
they are piecewise-linear functions.
\begin{figure}
  \includegraphics[width=\columnwidth]{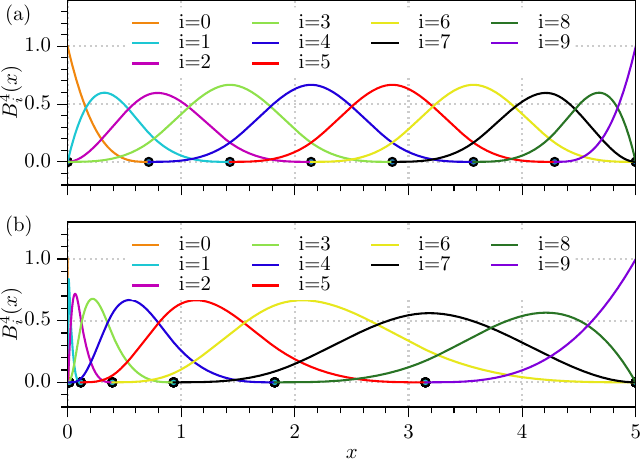}
  \caption{B-spline functions with (a) equidistant and (b)
    non-equidistant spacing of knots and additional ghost knots at the
    boundaries. When the first and last B-splines are removed, the
    functions satisfy vanishing boundary conditions.}
  \label{fig:Bsplines}
\end{figure}
Some B-splines with $k=4$ and different knots are illustrated in
Fig.~\ref{fig:Bsplines}.
Three important properties of B-splines should be mentioned.
Firstly, the B-spline of order $k$ on a grid of equidistant knots
approximates an analytical function with accuracy of about $h^{k}$,
where $h$ is the step size of the grid.
Thus, high-order B-splines can give a precise solution even with a
relatively small number of knots.
Secondly, one can choose the knots non-equidistantly and add ghost
knots at the boundaries in such a way as to have two B-splines equal to 1
in the two boundary points, while all other B-splines vanish in these
points (see Fig.~\ref{fig:Bsplines}).
Then, for example, zero boundary conditions can be easily implemented
by removing the first and the last B-splines from the basis.
Thirdly, the B-splines are nonorthogonal functions, and thus
Eq.~\eqref{eq:schrodinger_eq} turns into a generalized eigenvalue
problem
\begin{equation}
  \label{eq:GEVP}
  \sum_n H_{mn}c_n = E\sum_n O_{mn}c_n \, ,
\end{equation}
with $O_{mn}$ the overlap matrix between the basis functions.
However, the B-spline functions have minimal support, i.e., each B-spline
$B^{k}_{i}(x) = 0$ for $x \notin [t_{i},t_{i+k}]$, which significantly
reduces the number of integrations to calculate matrix elements.
This leads to the band structure of matrices or to
sparse matrices of the generalized eigenvalue problem~\eqref{eq:GEVP}
for one- or higher-dimensional problems, respectively.

When applying B-splines to the QW problem with the Hamiltonian
\eqref{eq:Hamiltonian_in_com_coordinates} it should be noted that the
QW potentials imply zero boundary conditions at $z_{\mathrm{e},\mathrm{h}}=\pm L/2$.
This can be achieved by a proper choice of the B-spline basis.
The boundary conditions in the $\rho$-direction are less obvious.
The wave function $\psi$ is finite but not necessarily zero at $\rho=0$.
To implement a zero boundary condition at $\rho=0$,
we use the substitution $\psi=\chi/\sqrt{\rho}$.
As a results of the substitution, the B-spline expansion of $\chi$ is
employed with the Hamiltonian
\begin{align}
  \label{eq:Hamiltonian_substitution}
	H &= E_{\mathrm{g}} - \frac{\hbar^2}{2\mu}\left(\frac{\partial^2}{\partial\rho^2}
	-\frac{m^2-1/4}{\rho^2}\right)
	- \frac{\hbar^2}{2m_{\mathrm{e}}}\frac{\partial^2}{\partial z_{\mathrm{e}}^2}
	- \frac{\hbar^2}{2m_{\mathrm{h}}}\frac{\partial^2}{\partial z_{\mathrm{h}}^2} \nonumber\\
	&- \frac{1}{4\pi\varepsilon_0\varepsilon}
	\frac{e^2}{\sqrt{\rho^2+(z_{\mathrm{e}}-z_{\mathrm{h}})^2}}
\end{align}
and the zero boundary conditions at $z_{\mathrm{e},\mathrm{h}}=\pm L/2$.
Note that now the QW potentials in Eq.~\eqref{eq:Hamiltonian_in_com_coordinates}
are ignored in Eq.~\eqref{eq:Hamiltonian_substitution}.
We artificially introduce an additional zero boundary condition at a
large value $\rho=\rho_{\max}$, thus making a box-like geometry in this direction.
If the parameter $\rho_{\max}$ is sufficiently large, the spectrum of
bound states below the scattering threshold is well approximated.
However, without application of further methods we obtain a discretized
continuum instead of the true resonances above the thresholds.

In our calculations, we use the expansion
\begin{equation}\label{BSplineExpansion}
	\chi(z_\mathrm{e},z_\mathrm{h},\rho) \approx \sum_{l=1}^{\tilde{N}_{z_\mathrm{e}}}\sum_{m=1}^{\tilde{N}_{z_\mathrm{e}}}\sum_{n=1}^{\tilde{N}_{z_\mathrm{h}}}c_{lmn} B_{l+1}^k(z_\mathrm{e})   B_{m+1}^k(z_\mathrm{h})  B_{n+1}^k(\rho)\,
\end{equation}
with B-splines of order $k=5$ and
equidistant knots in the $z$-directions, but non-equidistant knots in
the $\rho$-direction similarly as illustrated in Fig.~\ref{fig:Bsplines}(b).
In Eq.~\eqref{BSplineExpansion} $\tilde{N}_q=N_q+k-4$, where $N_q$
is the number of physical knots in the $q$ direction.
For all directions, $k-1$ ghost knots were inserted at the boundaries, to be 
able to define at least $k-1$ B-splines at each interval.
For the $\rho$-direction the $i$-th knot between the ghost knots is calculated via
\begin{equation}
	\rho_{i+k-1}=\left (  \frac{i-k}{N_\rho-1}\right)^3 \rho_{\max}\,\,.
\end{equation}
We used 30 physical knots for the $\rho$-direction
and 22 knots for each of two $z$-coordinates.
The matrix elements  were calculated numerically by application of a 15
point Gauss-Kronrod formula~\cite{Kahaner89}.
To speed up calculations, only matrix  elements that do not vanish due
to the finite support of the B-splines are calculated.
The generalized eigenvalue problem~\eqref{eq:GEVP} with symmetric,
sparse, and banded matrices, was solved by LAPACK routines~\cite{LAPACK}.

\subsection{Stabilization method}
The stabilization method allows for the calculation
of the complex resonance energies directly from the Hermitian
eigenvalue problem with the Hamiltonian~\eqref{eq:Hamiltonian_substitution}.
The idea is to analyze the real-valued spectrum of the
Hamiltonian~\eqref{eq:Hamiltonian_substitution} as a function of the
box size parameter $\rho_{\max}$.
This yields the so-called stabilization diagram, i.e., a graph containing
curves $E_j(\rho_{\max})$ with $j$ counting the (real-valued) eigenvalues.
For sufficiently large values of $\rho_{\max}$, the energies belonging to
resonant states stabilize, i.e., they are almost independent of
$\rho_{\max}$ with the exception of regions close to avoided crossings
with energies of the discretized continuum in the box.
The density of states $\varrho(E)$ is then extracted from the slopes of the
curves in  the stabilization diagram as~\cite{Mandelshtam1993,Mandelshtam1994}
\begin{equation}
  \label{eq:density}
  \varrho(E) = \frac{1}{\Delta\rho_{\max}} \sum_j \left | 
    \frac{\text{d}E_j(\rho'_{\max})}{\text{d}\rho'_{\max}} \right |^{-1}_{E_j(\rho'_{\max})=E}.
\end{equation}
In open systems the density of states is typically given as a
superposition of Lorentzians, and therefore, in the final step, the
energy positions $E_{\mathrm{res}}$ and widths $\Gamma$ of isolated
resonances can be extracted from the density
profile~\eqref{eq:density} by fitting
\begin{equation}
  \label{eq:lorentz}
  \varrho(E)\simeq \pi^{-1} \frac{\Gamma/2}{(E_{\mathrm{res}}-E)^2+\Gamma^2/4} \, .
\end{equation}

\subsection{Complex-coordinate-rotation method}
The basic idea of the complex-coordinate-rotation method is to do the
transformation $r \rightarrow r\,e^{i\theta}$ with $\theta>0$, which
yields a non-Hermitian Hamiltonian.
The energies of the bound states are invariant under this transformation,
while the energies of the continuum states are rotated into the lower
half of the complex plane by the angle $2\theta$.
Most interesting for our application is the impact on resonant states.
The complex eigenvalues, corresponding to the resonant states, are
located in the sector of the angle $2\theta$ between the lowest
rotated branch of the continuum and the real axis.
They are independent or only weakly dependent of the angle of the
rotation if the angle is large enough to contain the unknown eigenvalue.

For our problem the complex-coordinate rotation is restricted to
the $\rho$-coordinate because the system is open only in this direction.
The boundary conditions in the $z$-directions must not be changed by a
complex rotation.
The substitution $\displaystyle\rho \rightarrow \rho e^{i\theta}$
transforms the Hamiltonian~\eqref{eq:Hamiltonian_substitution} to
\begin{align}\label{eq:rotated_hamiltonian}
	H &= E_{\mathrm{g}} -e^{-2i\theta} \frac{\hbar^2}{2\mu}\left(\frac{\partial^2}{\partial\rho^2}
	-\frac{m^2-1/4}{\rho^2}\right)
	- \frac{\hbar^2}{2m_{\mathrm{e}}}\frac{\partial^2}{\partial z_{\mathrm{e}}^2} \nonumber\\
	&- \frac{\hbar^2}{2m_{\mathrm{h}}}\frac{\partial^2}{\partial z_{\mathrm{h}}^2}
	- \frac{1}{4\pi\varepsilon_0\varepsilon}
	\frac{e^2}{\sqrt{e^{2i\theta}\rho^2+(z_{\mathrm{e}}-z_{\mathrm{h}})^2}} \, .
\end{align}
This leads to the generalized eigenvalue problem~\eqref{eq:GEVP} with
non-Hermitian, complex symmetric matrices, which is efficiently
solved by using the ARPACK package~\cite{arpackuserguide}.
Note that the angle $\theta$ must be chosen appropriately
to obtain convergence of the results~\cite{Moiseyev1978}.

\section{Results and discussion}
\begin{figure}
  \includegraphics[width=\columnwidth]{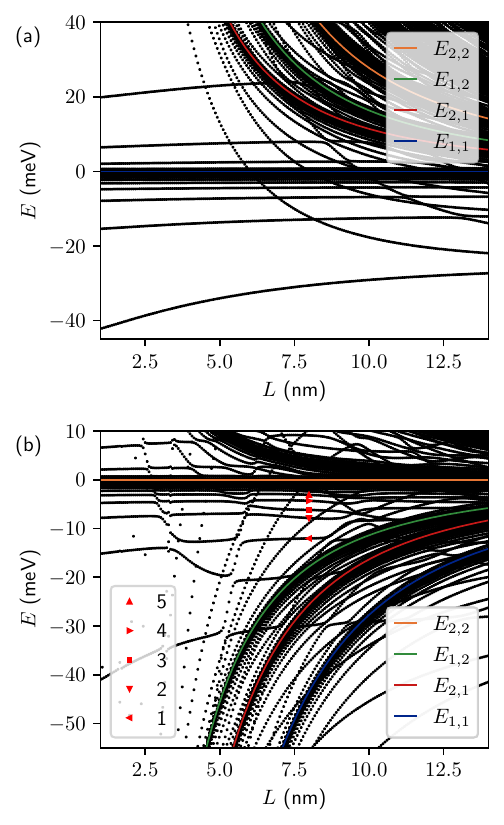}
  \caption{Spectrum of the
    Hamiltonian~\eqref{eq:Hamiltonian_substitution} with respect to
    (a) the threshold $E_{1,1}(L)$ and (b) the threshold $E_{2,2}(L)$ as a
    function of the QW width. Note that the energy scales
    differ from the energy scale in Fig.~\ref{fig:complex-rot}, where absolute
    values are given.
    The angular quantum number is $m=1$,
    and the box size parameter is $\rho_{\max}=500\,$nm.
    The five resonant states at $L=8\,$nm , which are investigated
    with the stabilization method and the complex-coordinate-rotation
    method are highlighted in (b) by red markers.}
  \label{fig:spectrum-2_2}
\end{figure}

In Fig.~\ref{fig:spectrum-2_2}(a) we present the 
real-valued energy eigenvalues for excitons in cuprous oxide QWs
described by the Hermitian Hamiltonian~\eqref{eq:Hamiltonian_substitution}
as a function of the width $L$ of the QW.
The computations have been performed for angular quantum number $m=1$
and for a fixed value of the box size parameter $\rho_{\max}=500\,$nm.
The first four thresholds $E_{i,j}(L)=E_{\mathrm{e}i}(L) + E_{\mathrm{h}j}(L)$
[see Eq.~\eqref{eq:thresholds}] are marked as colored lines.
For a better visualization of the bound states the energy of the first
threshold $E_{1,1}$ has been subtracted.
As explained above, only the bound states, i.e., the even parity states
below threshold $E_{1,1}$ and the odd parity states below the threshold
$E_{2,1}$ are converged states, which do not depend on the box size
$\rho_{\max}$.
The energy levels of the bound states exhibit
the crossover from the model of the narrow QW (2D Coulomb potential)
at low values of $L$ to the model of the bulk crystal (3D Coulomb
potential) at large $L$~\cite{Belov2019}.
When the QW width is increased, the thresholds gradually descent to
the lowest one.
As a result, the states associated to upper thresholds get down below
the lowest threshold and the number of bound states increases.
In contrast to the limiting cases, for intermediate QW widths the
  principal quantum number $n$ and the quantum-confinement ones
$i$ and $j$ lose their status as good quantum
  numbers. However, it remains possible to assign approximate quantum
  numbers to most of the states.
A detailed discussion of the bound states is given in Ref.~\cite{Belov2023}.
\begin{figure}
  \includegraphics[width=1\columnwidth]{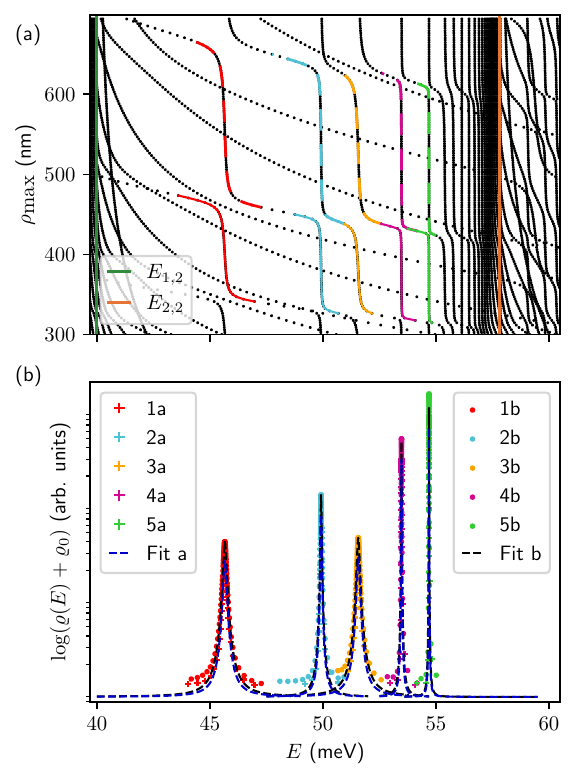}
  \caption{Application of the stabilization method for the QW width
    $L=\SI{8}{nm}$ and $m=1$ for the resonances highlighted in
      Fig.~\ref{fig:spectrum-2_2}(b). Note that here we show absolute
      resonance energies, which differ from the values in
      Fig.~\ref{fig:spectrum-2_2}(b), where the threshold energy
      $E_{2,2}(L)$ has been subtracted.  (a) Stabilization diagram in the region
    from $\rho_{\max}=300\,$nm to $\rho_{\max}=700\,$nm.  Segments
    belonging to the studied resonant states  are highlighted by the
    colored solid and dashed lines. They refer to the states 1-5 in
    Fig.~\ref{fig:spectrum-2_2}(b).
    (b) Logarithm of the density of states as a function of energy $E$.
    Colored peaks marked as `a' and `b' are obtained from the solid
    and dashed segments in (a), respectively.
    For a better visualization, $\varrho(E)$ is shifted upwards by a
    small value $\varrho_0$, so that $\log(\varrho)$ is bounded from below by a
    constant value. The crosses and the dots represent $\varrho(E)$
    calculated by Eq.~\eqref{eq:density}. The dashed lines represent
    the fits to the Lorentzian shape given in Eq.~\eqref{eq:lorentz}.}
  \label{fig:stab_meth}
\end{figure}
All other states above the respective thresholds are, in principle,
the unphysical discretized continuum states.
The continuum region becomes more pronounced in
Fig.~\ref{fig:spectrum-2_2}(b), where we show the same states as in (a)
but with respect to the threshold $E_{2,2}$.
We highlight five of the discretized continuum states with approximate
principal quantum numbers $n=3$ to $6$ (belonging to different thresholds)
at QW width $L=8\,$nm by red markers.
Such a QW width corresponds to the crossover region of QW widths, when
none of the limiting models is applicable.
On the one hand, this region is interesting due to the fact that
Rydberg excitons with principal quantum numbers $n\sim 3$ to $6$
are significantly disturbed compared to bulk excitons.
On the other hand, the Rydberg series related to the different
channels do not overlap too strongly to complicate the analysis of
spectra and the discussion.
A recent study demonstrated the feasibility of fabricating
nanostructures composed of Cu$_2$O at this scale~\cite{Awal2024}.
Since these states are in the continuum, the related resonances have
nonzero linewidths.
Note from Fig.~\ref{fig:spectrum-2_2}(b) that there is an additional
discretized continuum state a little bit below of the second marked
resonance state at $L=8$~nm.
It originates purely from the rigid-wall boundary conditions,
therefore we disregard this unphysical eigenvalue.

Although Fig.~\ref{fig:spectrum-2_2} provides a massive amount of
information on the dependence of Rydberg energies on the QW width,
it is not sufficient to determine the resonance linewidths.
To this end, we employed the two above-mentioned methods.
As a first technique, we apply the stabilization method to obtain the
energies and linewidths of these resonances from the real-valued
numerical data of energy levels in the continuum region.
To this aim, the stabilization diagram shown in Fig.~\ref{fig:stab_meth}(a)
has been calculated in the region from $\rho_{\max}=300\,$nm to
$\rho_{\max}=700\,$nm.
Clearly visible is the stabilization of several resonant states as
segments of approximately vertical lines separated by avoided
crossings.
For five selected resonant states, lower and upper
segments, labeled as `a' and `b' in what follows,
are highlighted by solid and dashed colored lines, respectively.
If these resonant states were bound states they would not depend on
the box-size parameter $\rho_{\max}$ and thus appear, without
undergoing any avoided crossings, as exactly vertical lines in the
stabilization diagram.
The deviations from exact vertical lines and in particular the avoided
crossings with some of the $\rho_{\max}$-dependent continuum states
indicate the coupling to the continuum, i.e., the stabilized states
are resonances with finite lifetimes.
Their slopes and avoided crossings allow for the computation of their
line shapes via Eq.~\eqref{eq:density}.
The density of states resulting from the application of
Eq.~\eqref{eq:density} to the highlighted
segments in Fig.~\ref{fig:stab_meth}(a)
is presented in Fig.~\ref{fig:stab_meth}(b) and exhibits
peaks with various widths and heights at the resonance positions.
These peaks are fitted by Eq.~\eqref{eq:lorentz} to a
  Lorentzian shape to obtain the resonance positions
$E_{\mathrm{res}}$ and linewidths $\Gamma$.
\begin{table}
  \caption{Complex resonance energies (in meV) of the five selected
    states with approximate principal quantum numbers $n=3$ to 6,
    which refer to the thresholds $E_{2,2}$ and $E_{3,1}$ obtained
    with the stabilization method and the complex-coordinate-rotation method.
    The assigned approximate quantum numbers of the QW eigenstates and
    the principal quantum numbers $n$ are listed in the first three
    columns. The linewidths of the resonances are given as
    $\Gamma=-2\,\mathrm{Im}\,E$. For the analyzed states, the
    stabilization method was applied to two different segments
    highlighted by solid and dashed colored curves in
    Fig.~\ref{fig:stab_meth} (see text), resulting in the values
    denoted by indices `a' and `b', respectively.}
  \label{tab:resonances}
  \begin{center}
    \begin{tabular}{cccc|cc|cc}
      &&&& \multicolumn{2}{c|}{Stabilization method} & \multicolumn{2}{c}{Complex rotation}\\
       $i$&$j$ & $n$&segment & Re$\,E$ & Im$\,E$ &  Re$\,E$ & Im$\,E$ \\
      \hline
      \multirow{2}{*}{2} & \multirow{2}{*}{2}& \multirow{2}{*}{3} &1a  &  45.6577 & -0.0815 &  \multirow{2}{*}{45.6607} &\multirow{2}{*}{-0.0798 } \\
       & & &1b & 45.6547 & -0.0812 &  & \\
        \multirow{2}{*}{2} & \multirow{2}{*}{2}  & \multirow{2}{*}{4} &2a& 49.9181 & -0.0242 &  \multirow{2}{*}{49.9189} &\multirow{2}{*}{-0.0239 } \\
       & & &2b  & 49.9164 & -0.0229 &  &  \\
      \multirow{2}{*}{3}  & \multirow{2}{*}{1} & \multirow{2}{*}{3} & 3a & 51.5645 & -0.0670 &  \multirow{2}{*}{51.5665} &\multirow{2}{*}{-0.0691} \\
      & & & 3b & 51.5613 & -0.0705 &  & \\
      \multirow{2}{*}{2} & \multirow{2}{*}{2}   & \multirow{2}{*}{5} & 4a& 53.4736 & -0.0060 &  \multirow{2}{*}{53.4738} &\multirow{2}{*}{-0.0060} \\
      & & & 4b  & 53.4738 & -0.0062 &   &  \\
       \multirow{2}{*}{2}  & \multirow{2}{*}{2} & \multirow{2}{*}{6} &5a&54.6914 & -0.0020 &   \multirow{2}{*}{54.6915 } &\multirow{2}{*}{-0.0020} \\
      &  & & 5b  & 54.6914 & -0.0019 &  & 
    \end{tabular}
  \end{center}
\end{table}
The corresponding values obtained by the stabilization method are
given in columns 5 and 6 of Table~\ref{tab:resonances}.
Note that the densities of states as well as Lorentzian fits labeled
as `a' and `b' are obtained respectively from the lower and upper
series of highlighted segments in the stabilization diagram
[Fig.~\ref{fig:stab_meth}(a)].
Both fits for a given resonance and, as a result, the corresponding
complex energies in Table~\ref{tab:resonances} are nearly identical.
This indicates that the sum in Eq.~\eqref{eq:density} is well
approximated by the first terms.

\begin{figure}
\includegraphics[width=\columnwidth]{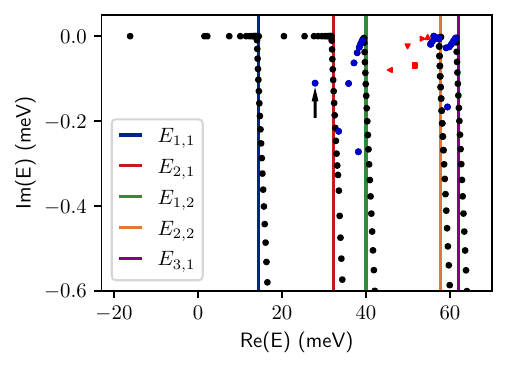}
  \caption{Eigenvalues of the complex-coordinate-rotated
    Hamiltonian~\eqref{eq:rotated_hamiltonian} for the QW width
    $L=\SI{8}{nm}$ and the angular momentum quantum number
    $m=1$.´Note that the resonance positions are shifted
    compared to Fig.~\ref{fig:spectrum-2_2}(b), since there no
    threshold energy has been subtracted.
    The first five threshold energies $E_{i,j}$ are shown as colored
    lines.  The five selected resonances, which have been analyzed by
    application of the stabilization method are marked by red  markers,
    while all other resonances are represented by blue dots.
    The even parity resonance below the threshold $E_{2,1}$ is marked by
    an arrow.}
  \label{fig:complex-rot}
\end{figure}

We now apply a second technique, and compare the data obtained by
the stabilization method with the results of the complex-coordinate-rotation
method, i.e. with the eigenvalues of the 
non-Hermitian 
Hamiltonian~\eqref{eq:rotated_hamiltonian}.
In our calculations, the angle of the coordinate rotation is $\theta=0.1$.
We note that the results presented in Fig.~\ref{fig:complex-rot} and
Table~\ref{tab:resonances} are fully converged to numerical accuracy
for rotation angles in the region $\theta\sim 0.1$ - $0.2$.
The resonance positions in the complex energy plane for the QW width
$L=8\,$nm and the angular momentum quantum number $m=1$ are shown
in Fig.~\ref{fig:complex-rot}.
The Rydberg bound states are located on the real axis with $\mathrm{Im}\,E=0$.
The scattering thresholds $E_{i,j}$ given in Eq.~\eqref{eq:thresholds}
are marked by colored vertical lines.
Clearly visible are the accumulation of even parity bound states with
$\pi_z=+1$ below the first scattering threshold $E_{1,1}$, the
accumulation of odd parity bound states with $\pi_z=-1$ below the
second (odd parity) scattering threshold $E_{2,1}$, and the discretized continuum
states related to the various scattering thresholds, which are rotated
into the lower half of the complex plane by the angle $2\theta$.
Although the Rydberg series of the odd parity bound states is in the
continuum background of the even parity states, the states of
different parity are uncoupled. As a result, the calculated
$\mathrm{Im}\,E$ of these odd parity bound states are zero to
numerical precision.
The remaining states are the resonance states with finite lifetimes,
i.e., $\mathrm{Im}\,E<0$; they are denoted by colored  markers.
The five selected resonances, which have already been studied by
application of the stabilization method, are marked by red  markers.
Their  resonance energies are listed in the two last columns of
Table~\ref{tab:resonances}.
One can see an excellent agreement between the results of the
stabilization method and the complex-coordinate-rotation method.

When the QW width is small, the quantum-confinement thresholds are
well separated.
When the QW width is increased, the thresholds gradually descent to
the lowest one.
As a result, the resonances of the current threshold get down below
the adjacent lower thresholds.
For QW width $L=8$~nm, such a penetration is shown by a small number
of resonances, that simplifies the current analysis.
For example, this  can clearly be
seen in Fig.~\ref{fig:complex-rot} by the even parity
resonance with the distinct complex eigenvalue $E=(27.8840-i0.1107)\,$meV
marked by the arrow, which is below the $E_{2,1}$ scattering threshold.

\section{Conclusions}
In QWs, multiple Rydberg series of electron-hole resonances appear
above the lowest scattering threshold due to the Coulomb coupling of
the upper quantum-confinement subbands to the continuum of lower ones.
These Rydberg series are well separated in case of the strong
confinement, therefore the approximate perturbative treatment of the
scattering problem is possible~\cite{Schmelcher2005}.
For arbitrary QW widths, in particular, for intermediate thicknesses,
a precise determination of the resonance parameters can only be done
using a numerical solution.
In our Paper, we identified the Rydberg series of electron-hole
resonances in cuprous oxide QW and accurately calculated their
energies and linewidths for the thickness, for which the resonances of
different thresholds do not significantly overlap.
To this end, the efficient numerical method based on the expansion of
the wave function of the original three-dimensional Schr\"{o}dinger
equation over a basis of B-splines was developed.
It made it possible to study the dependence of the numerical solution
on the size of the calculation domain in the broad range of the
parameter variance.
The lifetimes of the resonant states were calculated using the
stabilization method, allowing to derive the density of states and the
resonant parameters from the solution of the real-valued eigenvalue
problem.
The obtained numerical results are compared with the data calculated
by the complex-coordinate-rotation method.
The agreement of the results allows us to demonstrate the
applicability of both methods to study electron-hole resonances in QWs
of arbitrary widths.
In particular, we obtained that independent of the QW thickness, the
two lowest Rydberg series of states, below the second threshold, are
bound states.
The first series is associated to the even parity bound states and the
second one is associated to odd parity bound states in the even parity
continuum.
As a result, for the strong confinement, the nonvanishing broadening
is proper to the upper-lying resonance states.
However, with the crossover to the weak confinement, the partial
overlap of the Rydberg series is increased, resonances get down below,
and the analysis of the results becomes more complicated.

The presented methods to study the selected electron-hole resonances
open up the opportunity to investigate other features of resonant
states, e.g., the threshold effects of resonances by varying the QW
width.
Moreover, taking into account the full valence band structure of
cuprous oxide and using the eigenstates for the simulation of
absorption spectra will allow for detailed comparisons of our
theoretical results with future experimental measurements.

\acknowledgments
This work was supported by Deutsche Forschungsgemeinschaft (DFG)
through Grant No.\ MA 1639/16-1.
P.A.B.\ acknowledges financial support from Deutscher Akademi\-scher
Austauschdienst (DAAD) and the DFG Priority Programme 1929
\emph{``Giant interactions in Rydberg Systems''} (GiRyd), Grant No.\
SCHE 612/4-2.
We thank Stefan Scheel for fruitful discussions.

%

\end{document}